\documentclass[11pt,twocolumn,twoside,a4paper,amsmath,amssymb,aps,showkeys,showpacs]{revtex4}

\usepackage{amsfonts}
\usepackage{graphics} 
\usepackage{epsfig}
\usepackage{fancyheadings}

\begin{document}
\thispagestyle{myheadings}
\rhead[]{}
\lhead[]{}
\chead[SVD-2 Collaboration]{REGISTRATION OF NEUTRAL CHARMED MESONS PRODUCTION AND THEIR DECAYS 
IN pA-INTERACTIONS AT 70 GeV WITH SVD-2 SETUP}

\title{REGISTRATION OF NEUTRAL CHARMED MESONS PRODUCTION AND THEIR DECAYS 
IN pA-INTERACTIONS AT 70 GeV WITH SVD-2 SETUP}

\author{
{\bf (SVD-2 Collaboration)}\\
A. Aleev, V. Balandin, N. Furmanec, V. Kireev, G. Lanshikov, Yu. Petukhov, T. Topuria, A. Yukaev.}
\affiliation{%
Joint Institute for Nuclear Research, Dubna, Russia}%
\author{E. Ardashev, A. Afonin, M. Bogolyubsky, S. Golovnia, S. Gorokhov, V. Golovkin, A. Kholodenko, A. Kiriakov, V. Konstantinov, L. Kurchaninov, G. Mitrofanov, V. Petrov, A. Pleskach, V. Riadovikov*, V. Ronjin, V. Senko, N. Shalanda, M. Soldatov, Yu. Tsyupa, A. Vorobiev, V. Yakimchuk, V. Zapolsky.}
\affiliation{%
Institute for High Energy Physics, Protvino, Russia}%
\email{riadovikov@ihep.ru}
\author{S. Basiladze, S. Berezhnev, G. Bogdanova, V. Ejov, G. Ermakov, P. Ermolov, N. Grishin, Ya. Grishkevich, D. Karmanov, V. Kramarenko, A. Kubarovsky, A. Leflat, S. Lyutov, M. Merkin, V. Popov, D. Savrina, L. Tikhonova, A. Vischnevskaya, V. Volkov, A. Voronin, S. Zotkin, D. Zotkin, E. Zverev.}
\affiliation{%
D.V. Skobeltsyn Institute of Nuclear Physics, Lomonosov Moscow State University, Moscow, Russia}%

\begin{abstract}
The results of data handling for SERP-E-184 experiment obtained with 70 GeV proton beam irradiation of active target with carbon, silicon and lead plates are presented. Two-prongs neutral charmed  $D^0$ and $\bar D^0$ -mesons decays were selected. Signal / background ratio is (51$\pm$17) / (38$\pm$13). Registration efficiency for mesons was defined and evaluation for charm production cross section at threshold energy is presented: $\sigma(c\bar c) = 7.1\pm2.4(stat.)\pm1.4(syst.)$  $(\mu{b/nucleon})$.
\end{abstract}



\maketitle


\renewcommand{\thefootnote}{\roman{footnote}}


\section{Introduction}
\label{introduction}

The SVD-2 is universal experimental setup (Figure 1) \cite{ard01} with the following elements: Active Target (AT) with microstrip silicon plates (40 channels of electronics) and passive plates (C, Pb), Microstrip Vertex Detector (MVD) (8.5 thousand channels), Magnetic Spectrometer (MS) with Wire Proportional Chambers (MWPC) (near 18 thousand channels), Threshold Cherenkov Counter (TCC) (32 photo multipliers), Scintillation Hodoscopes (SH) (on 12 strips in vertical and horizontal planes) and the Detector of Gamma quanta (DEGA) with radiators from lead glass on 1344 channels. The data taken run was performed in the proton beam of IHEP accelerator with $E_p$=70 GeV. The total statistics of $5*10^7$ inelastic events has been obtained.

\begin{figure}
\includegraphics[scale=0.5]{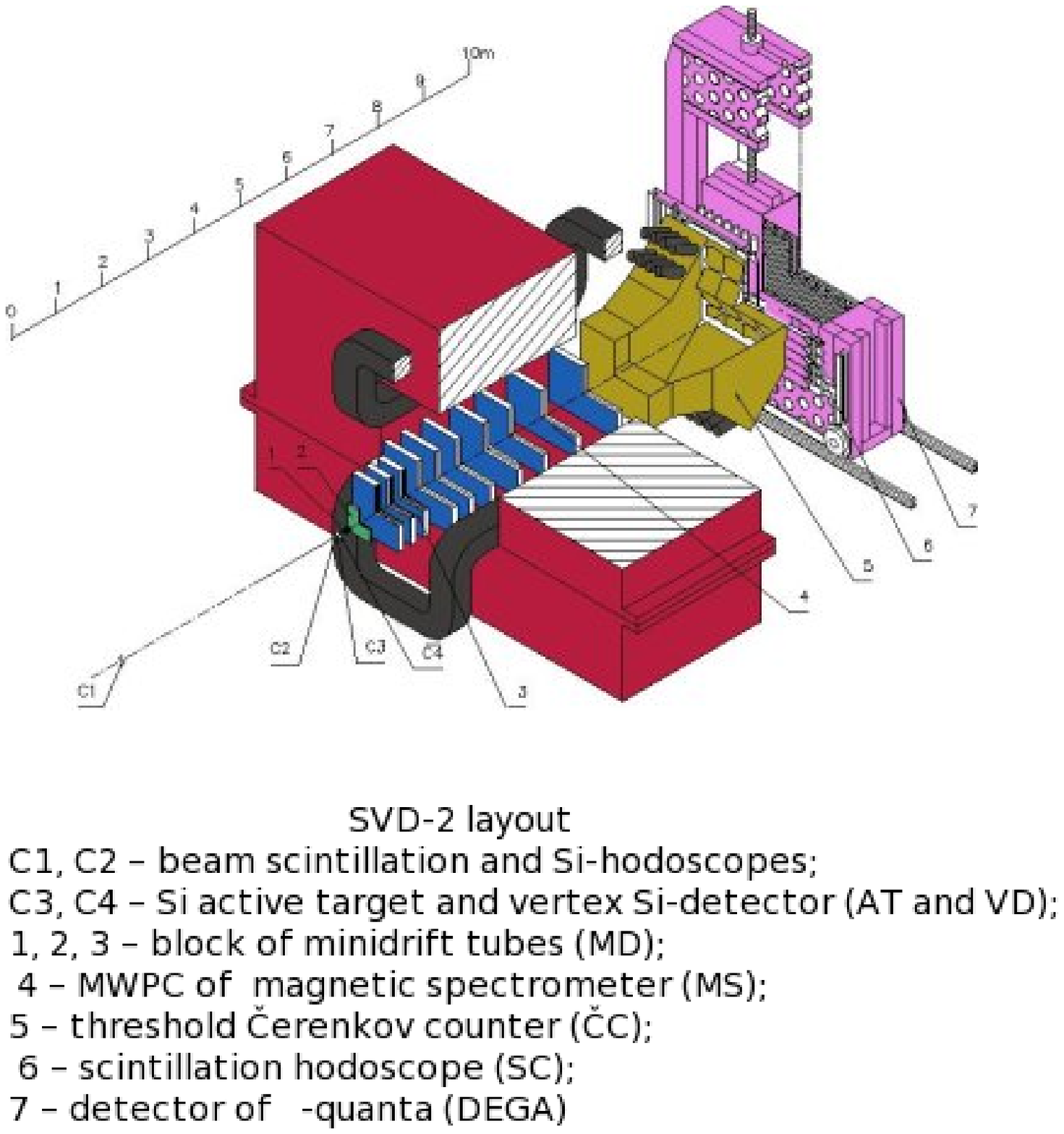}
\caption{\label{fig:wide}SVD-2 setup layout.}
\end{figure}

For last 20 years the obtained number of events with open charm production in a proton-nuclear interactions in experiments is much less than statistics of experiments with electron beams in which the basic properties of the charmed particles (mass, branching of decays, etc.) were studied. But the data of proton-nuclear experiments are important for studying of dynamics of charmed quarks production in collisions of nucleons and their hadronization mechanisms, for checking predictions of existing theoretical models.

Taking the cross section of inelastic pp-interaction ($\sigma_{in}(pp)=31.44$ mb at 70 Gev \cite{gor01}) and using the experimental fact, that the cross section of $c\bar c$-pair production in a nucleon-nuclear interactions linearly depends on atomic mass number of target, whereas "the usual" inelastic section is proportional $A^{0.7}$, it is possible to estimate the total number of events with charm in experiment E-184 for statistics of 52 million events with inelastic pA-interactions as
\begin{center}
$N(c\bar c) = N_0 * (\sigma(c\bar c)*A^1)/(\sigma_{in}(pp)*A^{0.7})$,
\end{center}
where $N_0$ - number of inelastic events in target with atomic mass number A, $\sigma(c\bar c)$ is accepted as 1 $\mu b$.

In Table I the experimental number of events in different plates of the target and number of expected events with charm are presented.

Using yields of particles and branching of their decays, it is possible to estimate number of events with different modes of charm particles decays (Table II).

The program FRITIOF 7.02 \cite{pi01} for the simulation of production processes has been used for D decays selection. Some parameters of model were adjusted with $D^0$ and $\bar D^0$ spectra measured in OPAL \cite{opal01} and CLEO \cite{huang01} experiments. Other ones were used by default. Simulation of charm decays products registration in SVD-2 setup was made with GEANT3.21 \cite{ge01}.

\begin{table*}
\caption{\label{tab:table3}Number of expected events with charm at $\sigma(c\bar c)=1 \mu b$}
\begin{ruledtabular}
\begin{tabular}{|c|c|c|c|c|c|c|}
The&Thickness&$N_{int}$&A&$A^{0.7}$&$N_0$&$N(c\bar c)$\\
target&(mkm)&(\%)&&&(million ev.)&(ev.)\\ \hline
C&540&21&12&5.7&10.92&732\\
Si&300*5=1500&55&28&10.3&28.60&2472\\
Pb&270&24&207&41.8&12.48&1966\\ \hline
TOTAL:&&100&&&52.00&5170\\
\end{tabular}
\end{ruledtabular}
\end{table*}

\begin{table*}
\caption{\label{tab:table3}Number of expected events with $D^0$ decays}
\begin{ruledtabular}
\begin{tabular}{|l|c|c|c|c|c|c|c|c|}
Decay&Branching&\multicolumn{2}{c|}{Carbon}&\multicolumn{2}{c|}{Silicon}&\multicolumn{2}{c|}{Lead}&$N_{tot}$\\ \cline{3-8}
&\cite{pdg01}&Yield&N&Yield&N&Yield&N&events\\
&&&events&&events&&events&\\ \hline
$D^0\rightarrow K^{-}\pi^{+}$&0.038&0.488&14&0.497&47&0.527&39&100\\
$\bar D^0\rightarrow K^{+}\pi^{-}$&0.038&0.590&16&0.585&55&0.578&43&114\\ \hline
TOTAL:&&&30&&102&&82&214\\
\end{tabular}
\end{ruledtabular}
\end{table*}

\section{Simulation of registration for events with $V^0$.}
\label{simulation}

The position of detectors and passive elements of the setup has been simulated using drawings with real metrological measurements and with results of alignment. The experimental map of magnetic field has been applied too \cite{bog01}. The probability of interaction in target plate has been calculated with taking account of its thickness and nuclear length. The sum of  seven probabilities are normalized to 1. Then the plate number with interaction was defined with random number generator [0,1]. The coordinate Z of the interaction along beam was defined with the center of the plate plus uniform displacement. Transverse coordinates (X, Y) of the vertex were pointed by the experimental beam profile. Particle kinematics in interaction point was given by FRITIOF7.02. There were 3 files with carbon, silicon and lead interactions. The decays of unstable particles were realized in GEANT. Appointed modes of decay were used for charmed mesons. During particle tracking in MVD the charge distribution along strips, the noise in each channel and the experimental amplitude cuts were taken into account. During the hit forming in MS experimental efficiencies of MWPC were used.

There are the following procedures of data handling system:
\begin{itemize}
\item Filtration of MVD data and the selection of events with second vertex near interaction point as the candidates for charmed events. The analysis in track parameters space \{a,b\} \cite{kir02} was used for that;
\item Geometrical reconstruction of tracks in MS and defining of momentum for charged particles;
\item The events analysis, including the kinematics of their.
\end{itemize}
The results of simulation were checked by means of comparison the distributions for MC and experimental data. There is a good consent in number of events for each plate of the AT. Multiplicity of charged particles in primary vertex (Figure 2a) and their momentums(Figure 2b) are in a good consent with model too.

\begin{figure}
\includegraphics[scale=0.3]{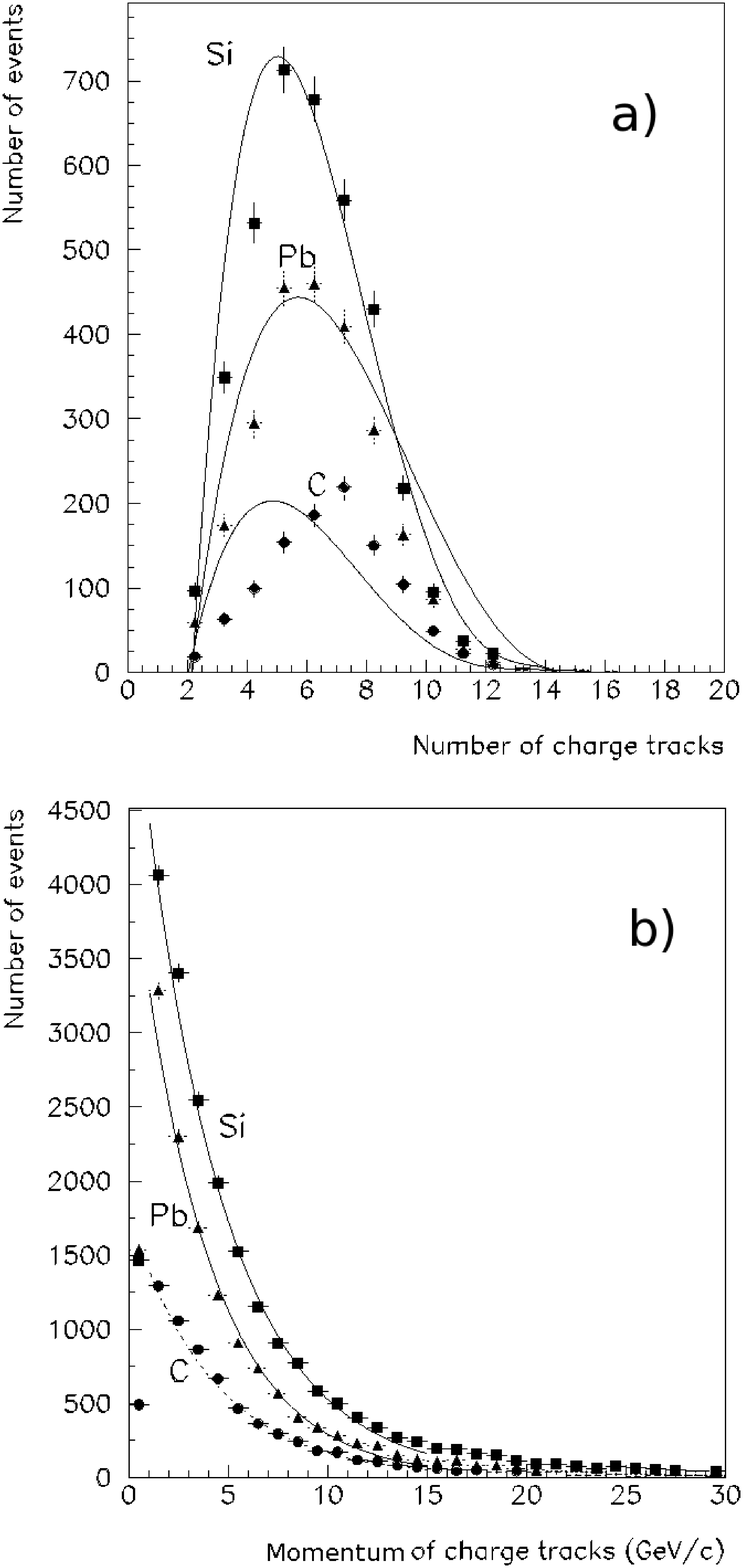}
\caption{\label{fig:wide}FRITIOF predictions (line) and experimental data ($\bullet - C, \blacksquare - Si, \blacktriangle - Pb$):\\
	a) the multiplicity of charged particles in primary vertex;\\
	b) the momentum of charged particles. }
\end{figure}

{\bf $K^0_s$-mesons registration.}

Simulation of $K^0_s$ registration in pA-interactions is a checking procedure for the searching of neutral D-mesons, because cross section of their production is known at our energy from MIRABELL data at 69 GeV/c \cite{am01}. There were 540000 MC events with 124000 $K^0_s$ and 85000 of them were with ($\pi^{+}\pi^{-}$)  decays. Next selection criteria have been applied for these events:
\begin{itemize}
\item the number of charged tracks $>$ 3;
\item the distance between primary vertex and $K^0_s$ decay $>$ 0.5 mm;
\item Z-coordinate of $K^0_s$ decay $<$ 35 mm;
\item tracks from $K^0_s$ decay should cross the last plane of MVD.
\end{itemize}
Only 2674 (=3\%) kaons  with ($\pi^{+}\pi^{-}$) decays are satisfied these conditions.

The criteria of $V^0$ events selection have been optimized at the simulation. They are the following: 
\begin{itemize}
\item the track from $V^0$ should have the impact parameter to primary vertex, $b/\sigma_b>2$;
\item two tracks from $V^0$ are crossing in the same space point, $\chi^2<4$ \cite{kir01};
\item the V0 vertex is separated from primary one, $(Z_2-Z_1)/sqrt(\sigma_1^2+\sigma_2^2)>3$;
\item the $V^0$ is coming through modified Podolanski-Armanteros criteria \cite{vor01}.
\end{itemize}
Fitting effective mass spectrum for ($\pi^{+}\pi^{-}$) system of MC events with Lorenz function for the signal and 2n-polynom for background we get $K^0_s$ mass = 497.5 MeV (in PDG – 497.6 МэВ), the width of signal =7.4 MeV. The efficiency for events with $K^0_s$ is 0.36\%.

{\bf $D^0$ and $\bar D^0$ registration.}

To optimize reconstruction procedure for two-prong decays ($K \pi$) of neutral D-mesons 100 thousand MC events with charm have been used. Number of $D^0$ was equal 51133, that correspond to yield of $D^0$. The simulation allows to estimate the efficiency of D-mesons registration (it is 7.2\%) and to optimize criteria of events selection. In Figure 3 it is shown as efficiency of registration depends on momentum and $X_{f}$ of mesons. For MC events $D^0$ mass is 1864 MeV and width of signal is 33 MeV.
The registration of two-prong decays of neutral anti D-mesons has been simulated too (Figure 4). In the aperture one gets 23\% $\bar D^0$ only because they fly back in c.m.s. mainly. Total efficiency of the registration is 2.7\%. For MC events $\bar D^0$ mass is 1866 MeV and width of signal is 36 MeV.

\begin{figure}
\includegraphics[scale=0.1]{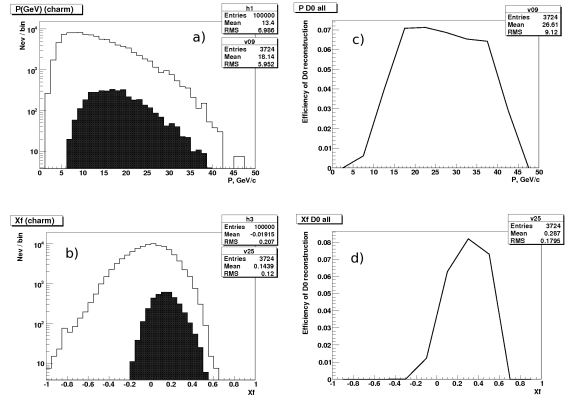}
\caption{\label{fig:wide} The momentum (a), $X_{f}$ (b)  for all and reconstructed $D^0$ (shaded); the efficiency of $D^0$ registration depending on momentum (c) and $X_{f}$ (d).}
\end{figure}

\begin{figure}
\includegraphics[scale=0.1]{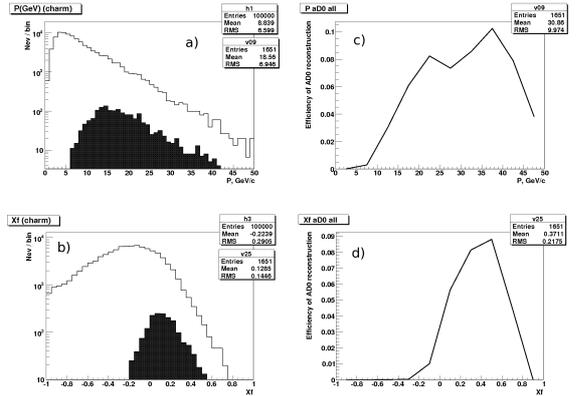}
\caption{\label{fig:wide} The momentum (a), $X_{f}$ (b)  for all and reconstructed $\bar D^0$ (shaded); the efficiency of $\bar D^0$ registration depending on momentum (c) and $X_{f}$ (d).}
\end{figure}

\section{Experimental data.}
\label{experiment}
{\bf Registration of $K^0_s$-mesons.}

To compare the results of simulation with experimental data, 1115091 experimental events with interaction in the active target have been processed by the same procedures. The fitted signal gives value of 498.6 MeV for $K^0_s$ mass, width of signal is 12 MeV, number of events in the signal is equal 222$\pm$20. In Figure 5 momentum (a) and $X_{f}$ (b) of reconstructed $K^0_s$ into strip of mass spectrum signal after leveling of histograms on the number of entries are compared for MC and experimental events.

\begin{figure}
\includegraphics[scale=0.17]{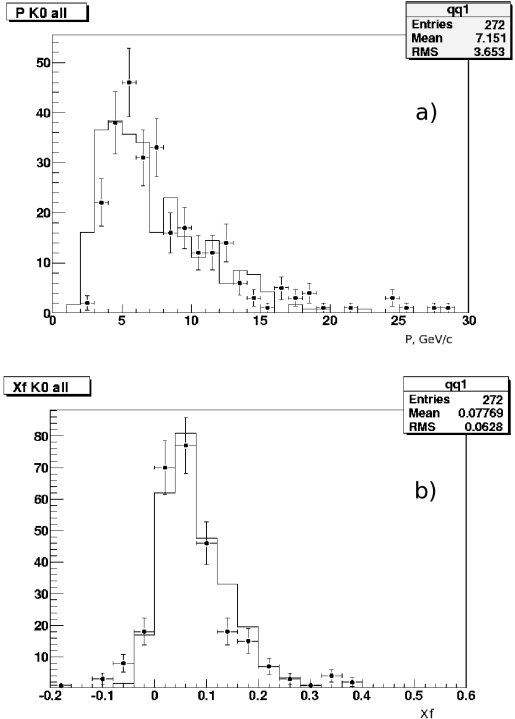}
\caption{\label{fig:wide}The momentum (a) and $X_{f}$ (b) of reconstructed $K^0_s$ for MC events (line) and for experimental events (points).}
\end{figure}

{\bf Selection of events with feasible decays of $D^0$-mesons.}

To select the events with possible two-prong decays ($K \pi$) of neutral D-mesons next basic criteria were applied:

\begin{enumerate}
\item The distance between primary vertex and $V^0$ decay should be more 0.5 mm;
\item The tracks from $V^0$ decay should have the impact parameter to primary vertex, but $V^0$ track should be pointed to it;
\item The effective mass of system ($K \pi$) should be in the range $\pm$0.5 GeV from $D^0$ mass (=1.865 GeV);
\item The momentum of system ($K \pi$) should be more 10 GeV/c;
\item $P_{t}$ of decay particle should be more 0.3 GeV/c to moving ($K \pi$) system according to Podolanski-Armanteros criteria and to suppression condition for background from kaon and $\Lambda^0$ (Figure 6);
\item The accepted hypothesis from two ($K^{-} \pi^{+}$) and ($K^{+} \pi^{-}$) modes has the mass nearest to $D^0$ mass;
\item $V^0$ should satisfy to selection criteria during visual inspection of events by physicist.
\end{enumerate}

\begin{figure}
\includegraphics[scale=0.4]{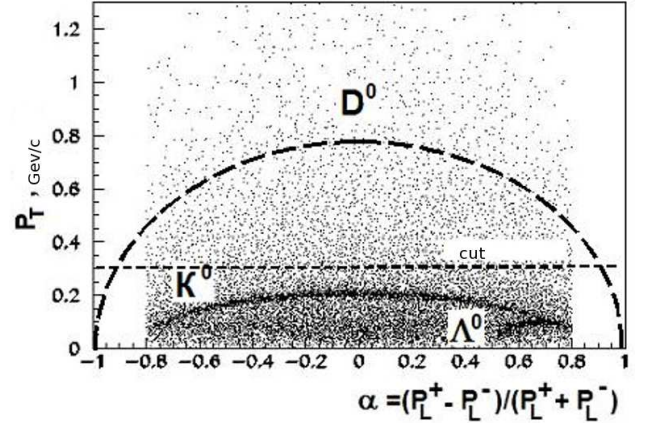}
\caption{\label{fig:wide}Podolanski-Armanteros plot for  $K^0_s$, $\Lambda^0$ and $D^0$.}
\end{figure}

It has been analyzed near 700 events. Principal causes for event rejection - the presence of background particles which were not reconstructed in space, but can belong to secondary vertex, the track from $V^0$ can belong to primary vertex, the possibility to accept $V^0$ instead of secondary interaction in the next plate of the target, etc. Additional background sources are:  the ignorance of real alignment and magnetic field, the intrinsic background from charm decay modes.

The effective mass spectra ($K \pi$) system before (a) and after (b) checking of events by physicist are presented in Figure 7. Because of small statistics, spectra for ($K^{-} \pi^{+}$) and ($K^{+} \pi^{-}$) systems were united in one spectrum. The area of mass, for which physical examination was carried out, has been limited from 1.7 to 2.0 GeV. In peak the signal/noise ratio is (51$\pm$17) / (38$\pm$13). Fit by straight line plus Gauss function for data after checking of events by physicist gives for $D^0$ mass 1861 MeV and width of signal $\sigma$=21 MeV.
Parameters of ($K \pi$) systems for the selected events in $M_{D}\pm 3\sigma$ range (momentum, $X_{f}$ and decay length) for MC events (line) and experiment (points) are presented in Figure 8. It is visible that the measured properties of experimental ($K \pi$) systems correspond to the properties of $D^0$-mesons from simulation.

\begin{figure}
\includegraphics[scale=0.35]{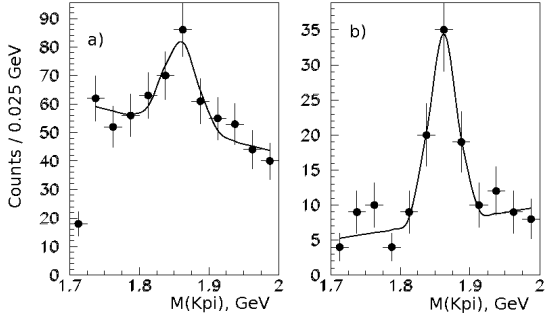}
\caption{\label{fig:wide}Effective mass spectrums ($K \pi$) system before (a) and after (b) checking of events by physicist.}
\end{figure}

\begin{figure}
\includegraphics[scale=0.4]{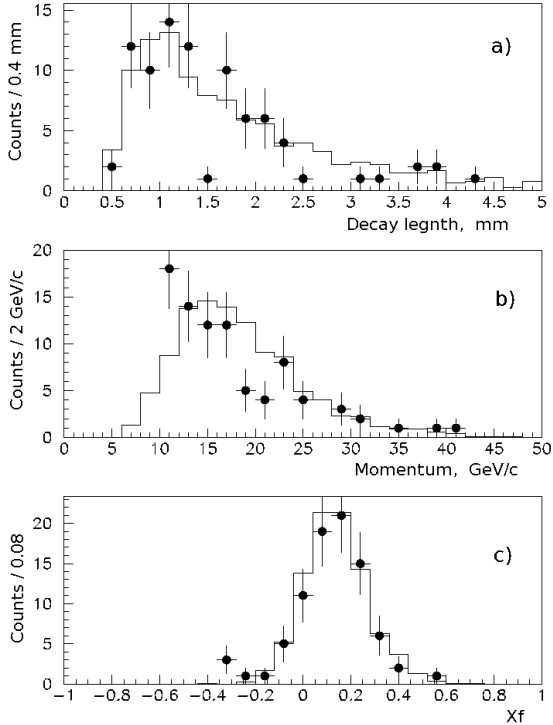}
\caption{\label{fig:wide}The decay length, the momentum and $X_{f}$ of ($K \pi$) system for MC and for experimental events.}
\end{figure}

\section{Estimations for charm production cross section.}
\label{estimations}

To estimate the cross section, except knowledge the number of events in the signal and the total statistics, the knowledge of other factors is necessary too: trigger factor, efficiencies of all procedures of data handling system.

The trigger factor, i. e. suppression of inelastic events registration during data acquisition has been estimated by comparison of charged particles multiplicity in primary vertex for MC and experimental inelastic events. The analysis of these distributions (Figure 9) gives $K_{trig}$= 0.51.

\begin{figure}
\includegraphics[scale=0.4]{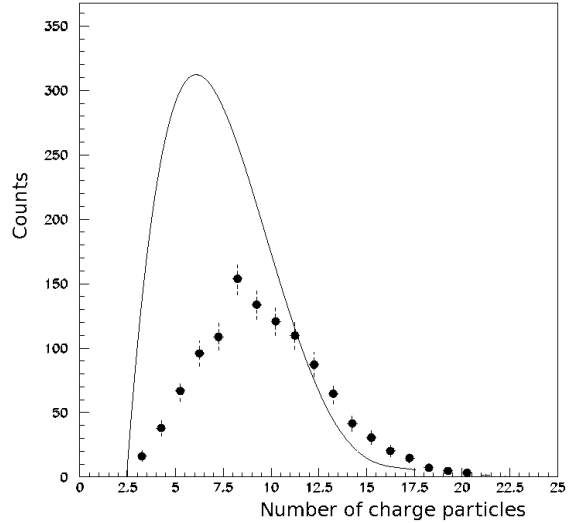}
\caption{\label{fig:wide}Charge particles multiplicity in the primary vertex for MC (line) and experimental (points) inelastic pA-interactions.}
\end{figure}

The fact is that the efficiency of the handling procedures decreases with experimental data. This factor is hardware factor and it can be estimated using of $K^0_s$ signal. $K^0_s$ production cross section at our energy for pp-interactions is known ($\sigma$=3430 $\mu b$) and A-dependence of cross section for pA-interactions is $A^{\alpha}$, where $\alpha$=0.78 \cite{abt01}. The efficiency for MC events with $K^0_s$ is 0.36\% and the number of decays $K^0_s\rightarrow \pi^{+} \pi^{-}$ may be
\begin{center}
$N(K^0_s)= 52*10^6*(3.43/31.44)*(A^{0.78}/A^{0.7})* 0.692*0.0036=19200$,
\end{center}
as  average atomic mass number of target nuclei is A=68. We have experimental signal from $K^0_s$ equal to  12000 decays only. Then $K_{hd}$=19200/12000=1.6.

The estimation of charm production cross section in our experiment has been made by two ways. We have: 
\begin{itemize}
\item Branching of decay $D^0\rightarrow K\pi$:  Br =0.038  \cite{pdg01} 
\item   Yield of $D^0$  = 49\%  \cite{kup01} 
\item   Number of events in signal:\\
$N_{reg}$ = 7(C)+22(Si)+22(Pb)=51 
\item   The registration efficiency (average on hypothesis):  $\epsilon$ = 0.036
\item   Predictions for the number of events with $\sigma(c \bar c)$=1 $\mu b$:
\begin{center}
$N_{pred}$= 30(C)+82(Si)+102(Pb)=214
\end{center}
\item  Trigger factor $K_{trig}$ = 0.51
\item  Hardware factor $K_{hd}$ = 1.6
\item  Integrated luminosity: $L_{int}$  = $N_{pA} /\epsilon /\sigma_{pN}$
\begin{center}
= [0.48(C)+1.27(Si)+0.55(Pb)] *$10^{33}$ =2.3*$10^{33}$ 
\end{center}
\end{itemize}

Method 1.

The first calculation of cross section is based on a prediction of the number of $D^0$ decays in our statistics from the assumption of linear dependence of charm cross section on $A$ and $A^{0.7}$ for cross section of inelastic interactions. Though, number of the allocated decays is small to leave a problem of averaging of atomic mass number, the cross section was calculated separately for each material of a target.
\begin{center}
$\sigma(c \bar c)$ = $K_{hd}*(N_{reg}/\epsilon) / (N_{pred} / K_{trig})$ =\\
= 5.3 (C); = 4.9 (Si); = 6.1 (Pb).
\end{center}
And the weighted average cross section is
\begin{center}
 $\sigma(c \bar c)$= 5.5 ($\mu b/nucleon$).
\end{center}

Method 2.

In the second calculation, integrated luminosity from the paper \cite{aleev01} is used. Integrated luminosity for each material of a target is defined, multiplying it by factor from Table I. As this luminosity corresponds to pA-events its value needs to be divided by $A^{0.7}$ that corresponds to A-dependence of cross section for inelastic interactions. We are obtaining $D^0$ production cross sections on a nucleus for different materials:
\begin{center}
$\sigma_{nuc} (D^0) = K_{hd} * N_{reg} / (Br*\epsilon) / (L_{int} / A^{0.7})$ =\\
=96.5 (C); = 209.5 (Si); = 1949.0 (Pb) ($\mu b$)
\end{center}

These values of cross sections show dependence on atomic mass number of target nuclei (A-dependence) with parameter $\alpha = 1.08 \pm 0.12$.

To compare these values to the cross section found above (Method 1) and to the data of other experiments, it is necessary to take into account A-dependence of charm production cross section ($\alpha$=1.0) from other experiments. Then we have $D^0$ production cross sections on a nucleon:
\begin{center}
$\sigma (D^0) = \sigma_{nuc} (D^0) / A$ = 8.0 (C); = 7.5 (Si); = 9.4 (Pb) ($\mu b/nucleon$)
\end{center}
To estimate value $\sigma(c \bar c)$ it is necessary to divide this value on production yield of neutral D-mesons which is equal 49\% by results of measurement in experiment \cite{kup01} and on number 2 since we consider the sum ($D^0 + \bar D^0$). Then we have
\begin{center}
$\sigma(c \bar c)$ = $\sigma (D^0)$ / 0.49 /2. = [8.2 (C); = 7.6 (Si); = 9.6 (Pb) ] = 8.7   ($\mu b/nucleon$)
\end{center}
We estimate uncertainties of measured cross section as 34\% (statistical) and 20\% (systematic). And the total error is 54\%.

\section{Conclusion.}
\label{conclusion}

The result of data handling in E-184 experiment for selected neutral D-mesons decays gives the estimation of charm production cross section in pA-interactions at near threshold energy 70 GeV:
\begin{center}
$\sigma(c \bar c)$ = 7.1 $\pm$ 2.4(stat.) $\pm$ 1.4(syst.) ($\mu b/nucleon$).
\end{center}
The dependence of cross section versus energy in c.m.s from \cite{shab01} is presented in Figure 10, including this result. The theoretical limits for cross sections are presented too.

The attempts for charm cross section estimation at threshold energy were made 20 years ago with BIS-2 IHEP setup, when carbon target was irradiated by neutrons at 40-70 GeV \cite{aleev02}. In kinematical range $X_{f}>0.5$ measured $D^0$ production cross section occurred much larger of theoretical predictions and is $\sigma (D^0)=28 \pm 14 (\mu b/nucleus)$. The cross section in our experiment is larger then theoretical estimations too. This fact demands more detailed study and much bigger stastics. The executed work is important, because the information on charm cross section at near threshold energy in pA-interactions is poor enough now.
\begin{figure}
\includegraphics[scale=0.4]{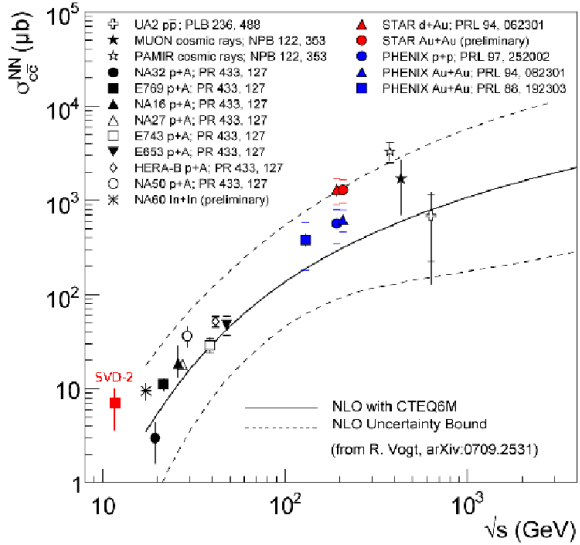}
\caption{\label{fig:wide}Experimental cross section of $D^0$ production in pA-interactions with the results of E-184 experiment.}
\end{figure}

\label{last}
\end{document}